\pdfoutput=1

\documentclass[conference]{IEEEtran}
\IEEEoverridecommandlockouts

\usepackage{cite}
\usepackage{amsmath,amssymb,amsfonts}
\usepackage{algorithmic}
\usepackage{graphicx}
\usepackage[numbers]{natbib}
\usepackage{textcomp}
\usepackage{xcolor}
\def\BibTeX{{\rm B\kern-.05em{\sc i\kern-.025em b}\kern-.08em
    T\kern-.1667em\lower.7ex\hbox{E}\kern-.125emX}}

\usepackage{tikz}

\newcommand\copyrighttext{%
  \footnotesize \textcopyright 2024 IEEE. Personal use of this material is permitted. Permission from IEEE must be obtained for all other uses, in any current or future media, including reprinting/republishing this material for advertising or promotional purposes, creating new collective works, for resale or redistribution to servers or lists, or reuse of any copyrighted component of this work in other works.}
\newcommand\copyrightnotice{%
\begin{tikzpicture}[remember picture,overlay]
\node[anchor=south,yshift=10pt] at (current page.south) {\fbox{\parbox{\dimexpr\textwidth-\fboxsep-\fboxrule\relax}{\copyrighttext}}};
\end{tikzpicture}%
}

\begin{document}

\title{Zero-Bit Transmission of Adaptive Pre- and De-emphasis Filters for Speech and Audio Coding\\}

\author{
\IEEEauthorblockN{Niloofar Omidi Piralideh, Philippe Gournay, Roch Lefebvre}
\IEEEauthorblockA{Department of Electrical and Computer Engineering\\
University of Sherbrooke\\
Québec, Canada \\
\{Niloofar.Omidi.Piralideh, Philippe.Gournay, Roch.Lefebvre\}@USherbrooke.ca}
}

\maketitle

\copyrightnotice

\begin{abstract}

This paper introduces a novel adaptation approach for first-order pre- and de-emphasis filters, an essential tool in many speech and audio codecs to increase coding efficiency and perceived quality. The proposed zero-bit self-adaptation approach differs from classical forward and backward adaptation approaches in that the de-emphasis coefficient is estimated at the receiver, from the decoded pre-emphasized signal. This eliminates the need to transmit information that arises from forward adaptation as well as the signal-filter lag that is inherent in backward adaptation. Evaluation results show that the de-emphasis coefficient can be estimated accurately from the decoded pre-emphasized signal and that the proposed zero-bit self-adaptation approach provides comparable subjective improvement to forward adaptation.

\end{abstract}
\vspace*{0.2cm}

\section{Pre-emphasis in speech and audio coding}

Lossy compression, also known as irreversible compression, achieves its goal by discarding certain information from the data to be compressed. When dealing with speech and audio signals, the primary method to discard information is quantization, an operation that consists in mapping a wide range of values to a more compact set of values. The resulting quantization error can often be modeled as additive white noise. This noise is obviously most noticeable at higher frequencies, where the amplitude spectrum of speech and audio signals is typically low \cite{rabiner2007introduction}.

To address this issue, Atal and Schroeder originally proposed a solution called pre- and de-emphasis \cite{atal1970adaptive}. Pre-emphasis consists in applying a linear filter to amplify high-frequency components of the input signal prior to its quantization or encoding. Following decoding, the signal needs to be de-emphasized using a matching linear filter so that it regains its original frequency balance.

The transfer function of a first-order pre-emphasis filter is:

\vspace*{-0.05cm}
\begin{equation}
A(z) = 1- \beta z^{-1}
\end{equation}
where $z^{-1}$ represents a delay element, and $0<\beta \leq1$ determines the strength of the pre-emphasis, a higher $\beta$ value resulting in stronger pre-emphasis. The value of $\beta $ varies between implementations and across different sampling rates. For instance, a constant $\beta $ of around 0.7 is commonly employed in various codecs such as EVS, AMR-WB, and G718 \cite{bessette2002adaptive}, \cite{3gpp-ts26.445}, \cite{rec2008g}, while OPUS uses a constant $\beta $ of 0.85 \cite{valin2016high}.

The main reasons for employing a pre-emphasis filter in a speech or audio coder are \cite{backstrom2017speech}, \cite{kabal2021ill}:
\begin{itemize}
\item To reduce the temporal and spectral dynamic range of the signal to be coded. This makes it easier to develop a fixed-point implementation of the codec.
\item To provide better signal conditioning for subsequent processing steps. For example, pre-emphasis leads to autocorrelation matrices that are better conditioned when performing linear prediction analysis.

\item To achieve some degree of spectral noise shaping. This increases subjective quality at very little cost in bit rate and complexity.
\end{itemize}

Although a fixed pre-emphasis filter is valuable in many codecs, it suffers from certain obvious limitations. Its inherent lack of adaptability can result in sub-optimal performance when confronted with rapid changes in signal characteristics. Additionally, there is a potential risk of overemphasizing high frequencies, which may introduce distortion or accentuate noise for specific signals with high-frequency content.

\begin{figure}[t!]
\raggedleft
\includegraphics[width=9cm,,height=2.7cm]{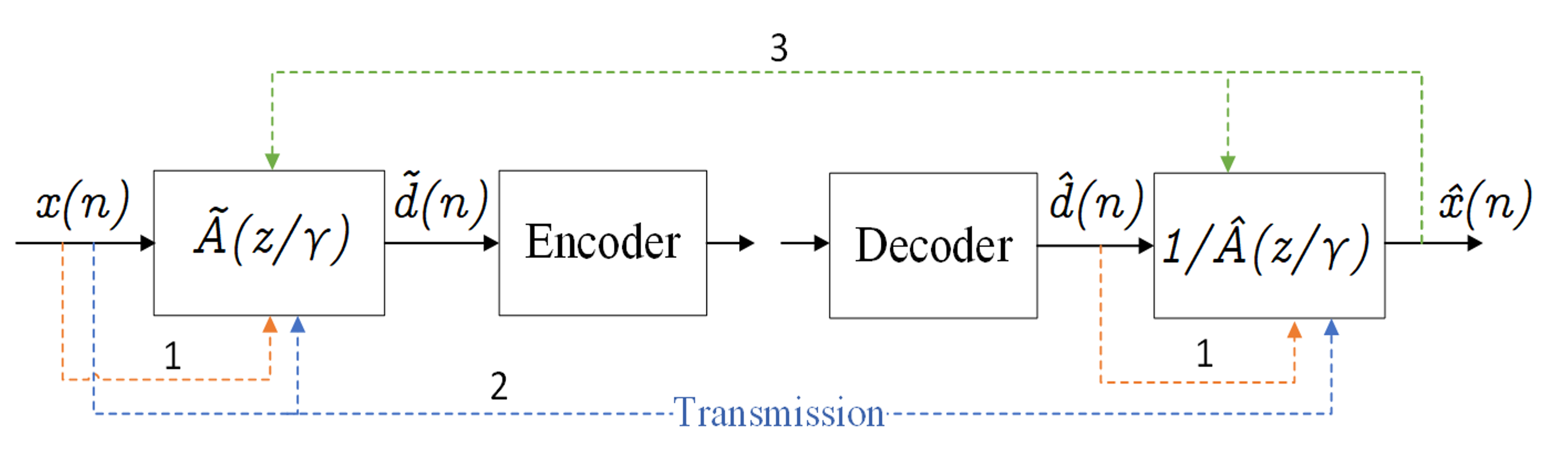}
\caption{A codec incorporating pre- and de-emphasis filters with (1) the proposed self-adaptation, (2) forward adaptation, and (3) backward adaptation.}
\label{fig:1}
\end{figure}

Unlike the fixed pre-emphasis filter, an adaptive pre-emphasis filter dynamically adjusts to the characteristics of the input signal, resulting in more effective signal conditioning and spectral noise shaping. There are two classical approaches for updating linear filters: forward and backward adaptation (see Fig. \ref{fig:1}). In forward adaptation, the filter is updated based on the input signal, which is unavailable to the decoder. Therefore, filter coefficients must be quantized and then transmitted to the decoder for de-emphasis operation.
In backward adaptation, the filter is updated based on the past synthesized signal, which is available to the decoder \cite{puri2000multimedia}.
Although backward adaptation achieves a lower bit rate than forward adaptation, the filter tends to lag the signal because its estimation can only be done from past synthesized samples. This lag is particularly detrimental when processing frames of a signal that exhibit rapid changes in statistical properties or spectral content.

Having identified the limitations of classical forward and backward adaptation approaches, we present a novel approach that efficiently addresses these limitations. The proposed approach completely eliminates the need to transmit additional information as well as the lag between filter and signal, offering a promising solution for efficient and effective adaptive pre- and de-emphasis filters in speech and audio coding.

\
\section{Self-adaptive pre- and de-emphasis filters}

The various adaptation approaches under study are illustrated in Fig. \ref{fig:1}. Throughout this paper, we will use the following notation:
\begin{itemize}
\item $A$ and $\alpha$: the real (generally unknown) first-order autoregressive (AR) model for signal $x(n)$ and the corresponding coefficient.
\item $\tilde{A}$ and $\tilde{\alpha}$: the first-order AR model and coefficient estimated at the encoder from the input signal, $x(n)$.
\item $\hat{A}$ and $\hat{\alpha}$: the first-order AR model and coefficient re-estimated at the decoder from the coded pre-emphasized signal, $\hat{d}(n)$.
\end{itemize}

Estimated and re-estimated models and coefficients are obviously identical in forward and backward adaptation, while they may be different to some degree in what we refer to as self-adaptation.

Given the above notations, the z-transform of the pre-emphasized signal $\tilde{d}(n)$ is given by:

\begin{equation}
\tilde{D}(z) =  X(z)\cdot(1-  \gamma \tilde{\alpha}z^{-1})
\label{eq:1}
\end{equation}
where $\gamma$ represents the filter weight (between zero and one) that controls the strength of the pre-emphasis. 
Subsequently, at the decoder, the z-transform of the reconstructed signal at the output of the de-emphasis filter is given by:

\begin{equation}
\hat{X}(z) = \hat{D}(z)\cdot\frac {1}{1- \gamma \hat{\alpha}z^{-1}}
\label{eq:1}
\end{equation}

\subsection{Mathematical derivation}

In this section, we derive the expression of the coefficient $\alpha$ of $A(z/\gamma)$ assuming $\gamma$ is known and using exclusively the pre-emphasized signal, $d(n)$. This is motivated by the fact that in our proposed self-adaptive approach, the filter coefficient at the decoder side is solely computed from the coded pre-emphasized signal.

To achieve this, we make the following simplifying hypothesis, as illustrated in Fig. \ref{fig:2}: the input signal, $x(n)$, is a first-order AR process (synthetic signal) generated by filtering white noise, $w(n)$, through a first-order AR filter, $1/A(z)$. Subsequently, applying the weighted filter $A(z/\gamma)$ to $x(n)$ yields a pre-emphasized signal, $d(n)$.

The input signal in Fig. \ref{fig:2} being a white noise, its autocorrelation coefficients are given by:

\begin{equation}
\label{eq:system1}
\begin{cases}
    \begin{aligned}
        & R_w(0) = E(w(n)^2) = \sigma^2 \\
        & R_w(k) = E[w(n) \cdot w(n-k)] = 0 \quad \text{for } k \neq 0
    \end{aligned}
\end{cases}
\end{equation}
where $E$ represents the mathematical expectation and $\sigma^2$ is the power of $w(n)$. 
In Fig. \ref{fig:2}, assuming a first-order autoregressive filter, the z-transform of the pre-emphasized signal, $d(n)$, can be calculated as:

\begin{equation}
D(z) = W(z) \cdot \frac{ 1 - \alpha \gamma z^{-1}}{ 1 - \alpha z^{-1}}
\label{eq:3}
\end{equation}
where $W(z)$ is the z-transform of input white noise. The second term in \eqref{eq:3} corresponds to an auto-regressive moving average model of order 1, denoted as ARMA(1,1), wherein the transfer function is a ratio of polynomials \cite{box2015time}.

Consequently, the time-domain expression of the pre-emphasized signal, $d(n)$, is:

\begin{equation}
d(n) =  \alpha d(n-1) + w(n) -  \alpha \gamma w(n-1)
\end{equation}
The first autocorrelation coefficient (lag 0) of the pre-emphasized signal is given by:

\begin{multline}
R_d(0) =  E[d(n)^2] =  \alpha E[d(n-1) \cdot d(n)] \\
+ E[w(n) \cdot d(n)]- \alpha \gamma E[w(n-1) \cdot d(n)] 
\label{eq:11}
\end{multline}
The second autocorrelation coefficient is expressed as:

\begin{multline}
R_d(1) =  E[d(n)\cdot d(n-1)] =  \alpha E[d(n-1)^2] \\
+ E[w(n) \cdot d(n-1)]- \alpha \gamma E[w(n-1) \cdot d(n-1)] 
\label{eq:13}
\end{multline}
To calculate the autocorrelation coefficients of $d(n)$, we need to first calculate the cross-correlation between $d(n)$ and $w(n)$: 

\begin{figure}
\centering
\includegraphics[width=6cm]{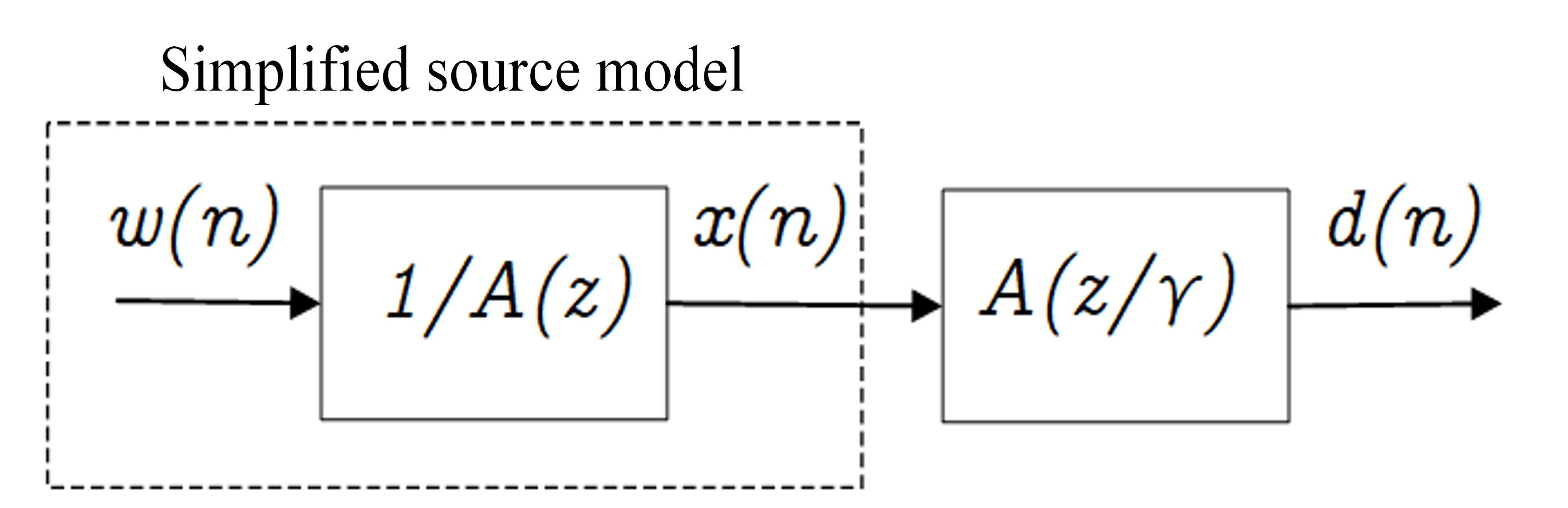}
\caption{Model assumption for pre-emphasized signal generation.}
\label{fig:2}
\end{figure}

\begin{multline}
E[d(n) \cdot w(n)] = \alpha E[d(n-1) \cdot w(n)] \\
+ E[w(n)^2] - \alpha \gamma E[w(n-1) \cdot w(n)]
\end{multline}
Given \eqref{eq:system1}, both $\alpha E[d(n-1) \cdot w(n)]$ and $\alpha \gamma E[w(n-1) \cdot w(n)]$ are equal to zero, and we have $E[w(n)^2] = \sigma^2$, therefore:

\begin{equation}
E[d(n) \cdot w(n)] = \sigma^2
\label{eq:7}
\end{equation}
Similarly, the cross-correlation between the pre-emphasized signal at time \(n\) and the white noise at the previous time \(n-1\) can be expressed as:

\begin{equation}
E[d(n) \cdot w(n-1)] = \alpha \sigma^2 - \alpha \gamma  \sigma^2
\label{eq:9}
\end{equation}
\\
Substituting \eqref{eq:9} and \eqref{eq:7} into \eqref{eq:11} yields:

\begin{equation}
R_d(0) = \alpha R_d(1) + \sigma^2 - \alpha \gamma (\alpha \sigma^2 - \alpha \gamma \sigma^2)
\label{eq:12}
\end{equation}
Here, substituting \eqref{eq:7} into \eqref{eq:13} yields:

\begin{equation}
R_d(1) = \alpha R_d(0) - \alpha \gamma\sigma^2 
\label{eq:14}
\end{equation}
Given that we have access to the coded pre-emphasized signal, $\hat{d}(n)$, at the decoder side, the values of $R_d(0)$ and $R_d(1)$ in \eqref{eq:12} and \eqref{eq:14} can be estimated.  
Consequently, the only variable yet to be determined is $\alpha$, which can be calculated by solving the system of equations formed by \eqref{eq:12} and \eqref{eq:14}:

\begin{equation}
\label{eq:system15}
\begin{cases}
    \begin{aligned}
        & R_d(0) - \alpha R_d(1) = \sigma^2 (1-\alpha^2\gamma +\alpha^2 \gamma^2) \\
        &  R_d(1) - \alpha R_d(0)  =  \sigma^2 (-\alpha \gamma)
    \end{aligned}
\end{cases}
\end{equation}
We can solve for \(\sigma^2\) from the second equation:

\begin{equation}
\sigma^2 = \frac {\alpha R_d(0) - R_d(1) }{\alpha \gamma}
\label{eq:16}
\end{equation}
Substituting this expression into the first equation of the system \eqref{eq:system15}, we obtain:

\begin{multline}
\alpha \gamma  R_d(0) - \alpha^2 \gamma  R_d(1)  = \\(\alpha R_d(0) - R_d(1) )(1-\alpha^2\gamma +\alpha^2 \gamma^2)
\label{eq:17}
\end{multline}
After some reorganization, we obtain the following equation that relates $\gamma$, $\alpha$ and $\frac{R_d(1)}{R_d(0)}$:

\begin{multline}
[\gamma(1-\gamma)] \alpha^3 + [\gamma \frac{R_d(1)}{R_d(0)} (\gamma-2)]  \alpha^2 \\ + [\gamma (\gamma-1)] \alpha + \frac{R_d(1)}{R_d(0)} = 0
\label{eq:18}
\end{multline}
The value of $\alpha$ is between -1 and 1, and solving (\ref{eq:18}) yields only one valid solution within this range among its three possible roots.

\vspace*{0.3cm}
\subsection{Application and tabulated numerical solution}
\begin{figure}
\centering
\includegraphics[width=9cm,height=7cm]{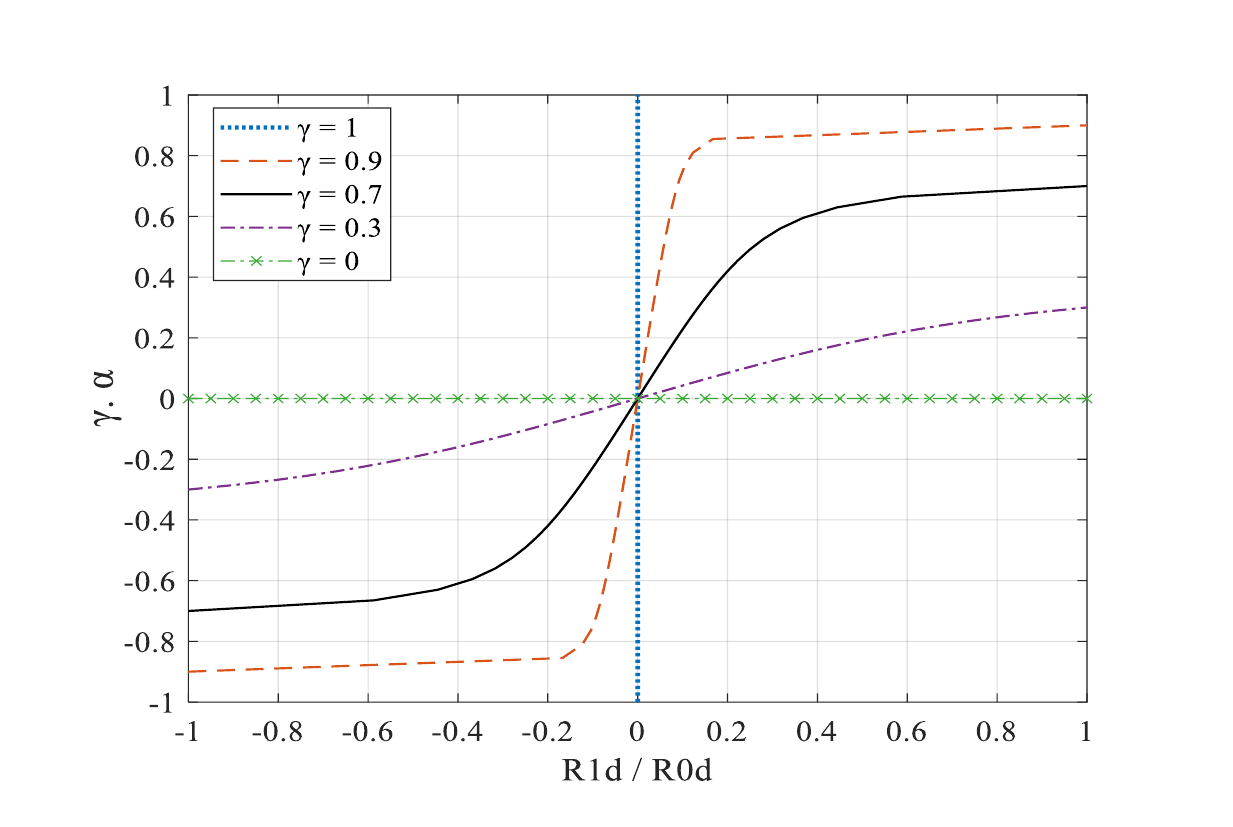}
\vspace*{-0.7cm}
\caption{Ideal de-emphasis coefficient $\gamma\cdot \alpha$ as a function of $\frac{R_d(1)}{R_d(0)}$ for different values of $\gamma$.}
\label{fig:3}
\end{figure}

\vspace*{0.2cm}
Fig. \ref{fig:3} shows the ideal de-emphasis coefficient $\gamma\cdot \alpha$, obtained by solving \eqref{eq:18} for $\alpha$, as a function of $\frac{R_d(1)}{R_d(0)}$ and for different values of $\gamma$. 

In practice, the pre-emphasis coefficient $\gamma\cdot\tilde{\alpha}$ is computed at the encoder side, with $\tilde{\alpha}$ being the least squares estimate of $\alpha$ given by $\frac{R_x(1)}{R_x(0)}$. The de-emphasis coefficient $\gamma\cdot\hat{\alpha}$ is computed at the decoder side, using $\frac{R_{\hat{d}}(1)}{R_{\hat{d}}(0)}$ and \eqref{eq:18}. Because of the simplifying hypothesis and the encoding-decoding operations in Fig. \ref{fig:1}, there is no guarantee that both coefficients will be strictly equal. This will be discussed in Section III.

Instead of solving \eqref{eq:18}, the curve corresponding to the chosen $\gamma$ can be tabulated and used to get $\hat{\alpha}$ from $\frac{R_{\hat{d}}(1)}{R_{\hat{d}}(0)}$  at minimal cost in computational complexity.

Fig. \ref{fig:3} also allows us to anticipate the performance of the proposed approach in various scenarios. For example, for values close to zero, a small variation in the estimate of $\frac{R_d(1)}{R_d(0)}$ can lead to a large variation in the estimate of $\gamma\cdot \alpha$. This will happen for example when $\hat{d}(n)$ is nearly white, either because the input signal $x(n)$ is itself nearly white ($\alpha$ close to 0) or because $\gamma$ is close to one (strong pre-emphasis). The proposed approach will be more accurate when $\hat{d}(n)$ is not nearly white.

\vspace*{0.18cm}
\section{Experiments and results}

\subsection{Simulations on synthetic first-order AR signals}

To evaluate the accuracy of the de-emphasis coefficient estimation on synthetic first-order AR signals (Fig. 2), we used Monte Carlo simulations \cite{mooney1997monte}. We varied the true $\alpha$ value from -0.98 to 0.98 by steps of 0.1. For each $\alpha$ value, we generated 30000 frames of 1440 samples of first-order AR signal, $x(n)$. Then, we computed $\tilde{\alpha}$ from $x(n)$ using the least-squares estimate $\frac{R_x(1)}{R_x(0)}$ and $\hat{\alpha}$ from $d(n)$ using \eqref{eq:18}. A 1440-sample Hanning window was used for autocorrelation computations. To achieve a good compromise between strength of the pre-emphasis and accuracy of the re-estimation, we set the value of $\gamma$ to 0.7. This value was found to give consistently good results in both objective and subjective evaluations. Notably, it aligns closely with the value used by many codecs.

\begin{figure}[t!]
  \centering
  \includegraphics[width=9cm]{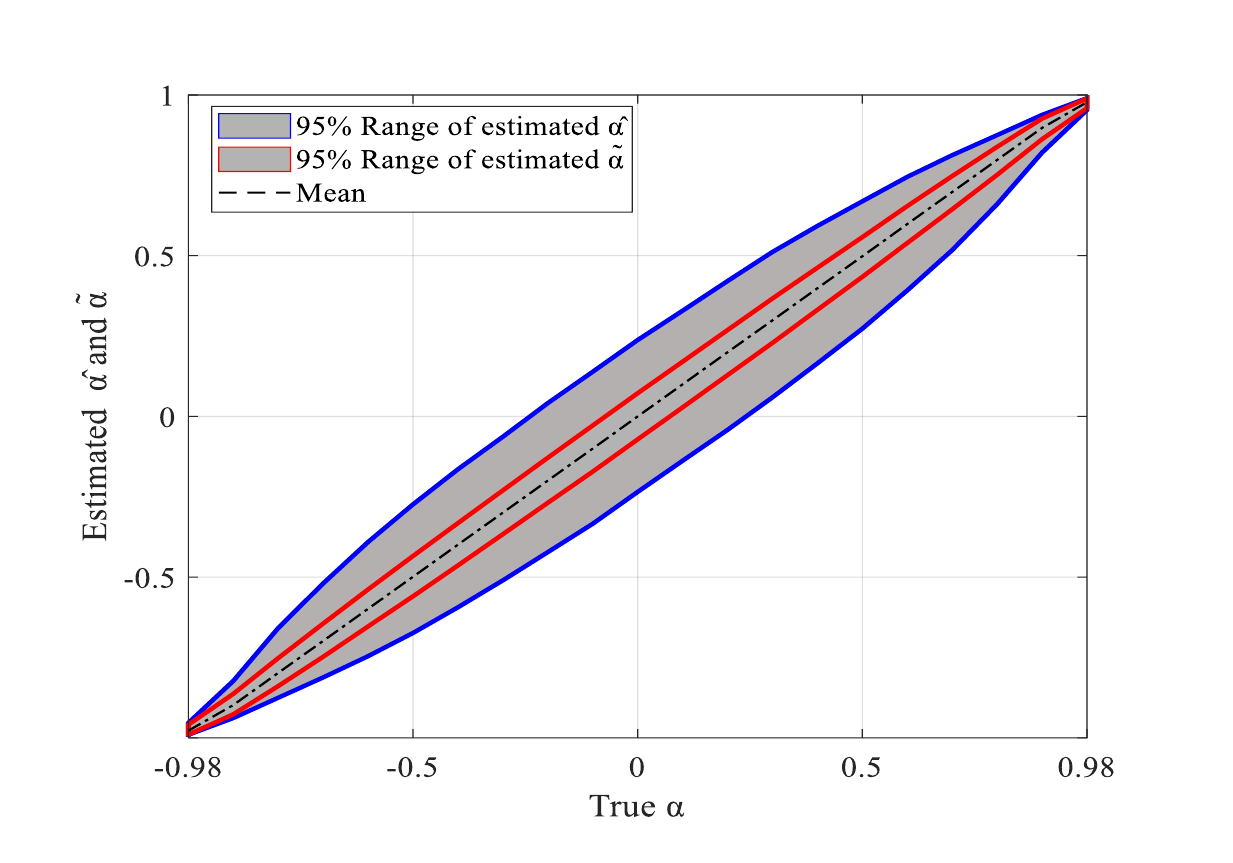}
  \caption{95\% range of estimated $\tilde{\alpha}$ and $\hat{\alpha}$ values, when $\gamma = 0.7$ .}
  \label{fig:4}
\end{figure}

Fig. \ref{fig:4} shows the 95\% central intervals of the estimated $\tilde{\alpha}$ and re-estimated $\hat{\alpha}$ relative to the true $\alpha$. It is possible to see that the estimates and re-estimates are more accurate for values of $\alpha$ close to one than for values close to zero.

\begin{figure}[htbp!]
  \includegraphics[width=9cm]{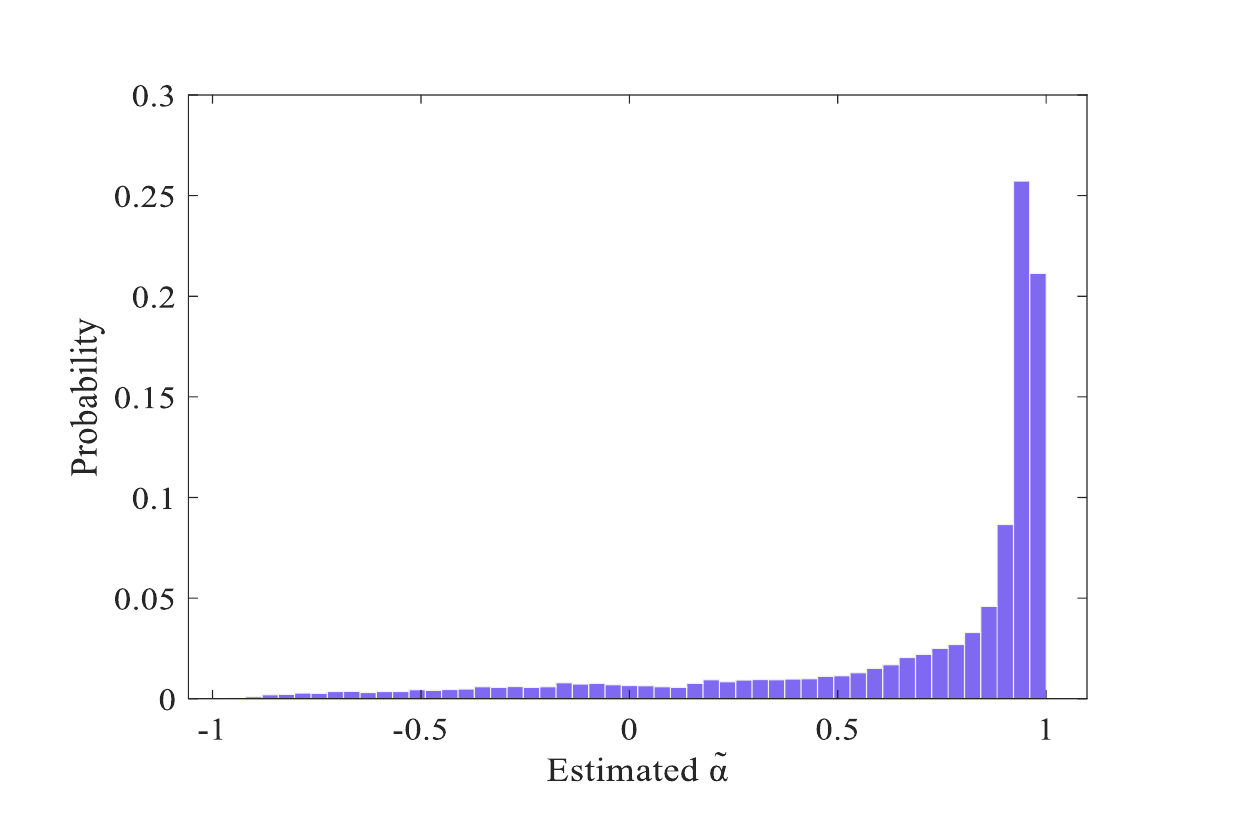}%
  \caption{Histogram of $\Tilde{\alpha}$ values extracted from a 16 kHz speech recording.}
  \label{fig:5}
\end{figure}

Fig. \ref{fig:5} shows the distribution of $\tilde{\alpha}$ extracted from a 5-minute recording of real speech signals (16 kHz sampling rate). It can be seen that speech signals often have $\tilde{\alpha}$ values close to one. This indicates that our approach should provide accurately re-estimated de-emphasis coefficients for most real speech signals.

\subsection{Impact of coding on real speech signals}

We evaluated the accuracy and invertibility of self-adaptive pre- and de-emphasis filters for speech signals in the presence of quantization noise. A 5-minute, 16 kHz speech sample was encoded using a PCM quantizer with additional self-adaptive and forward-adaptive pre- and de-emphasis filters. The frame duration was set to 10 ms (160 samples) and autocorrelations were computed using a 30-ms Hanning window with a 10 ms look-ahead. These parameters are typical in speech coding \cite{o1999speech}. The step size for the PCM quantizer was chosen to achieve the highest possible SNR for the 5-minute speech file.

The forward-adaptive approach utilized an unquantized coefficient. This configuration is used for comparison because, though it is not usable in practice, it theoretically outperforms both forward with quantization and backward adaptation.

\begin{figure}[t!]
  \includegraphics[width=9cm]{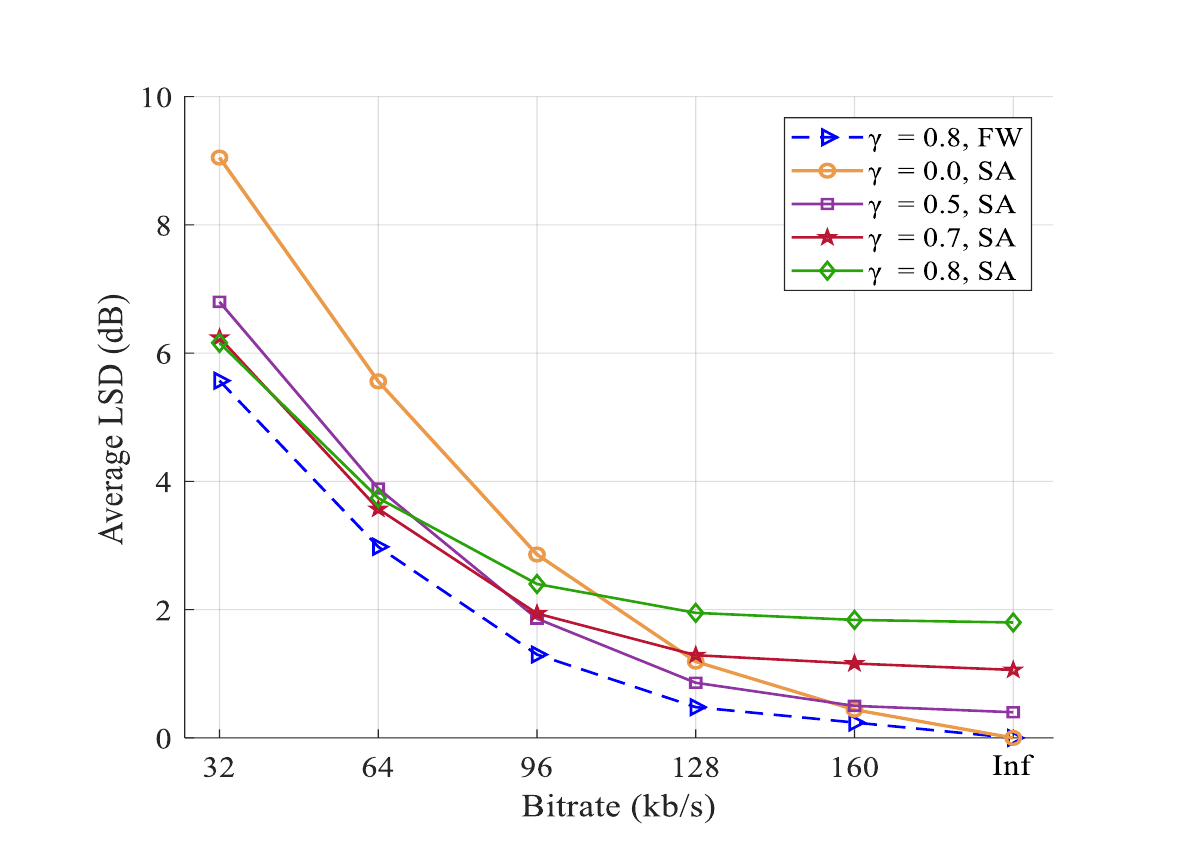}
  \caption{Average log spectral distortion (LSD) as a function of bit rate. Lower LSD is better.}
  \label{fig:6}
\end{figure}

To compare the performances of self- and forward adaptation across various bit rates, we computed the log spectral distortion (LSD) between the input signal, $x(n)$, and the de-emphasized output signal, $\hat{x}(n)$. The results are presented in Fig. \ref{fig:6}. In the case of self-adaptation, the LSD is plotted for various $\gamma$ values. In the forward-adaptive case, the LSD for a fixed $\gamma$ of 0.8 (yielding the highest subjective quality) is shown. The key observation from Fig. \ref{fig:6} is that the $\gamma$ value that minimizes spectral distortion depends on the bit rate. At higher rates, smaller $\gamma$ values lead to lower LSD, while at lower rates, larger $\gamma$ values result in lower LSD and better overall quality. Additionally, it is noteworthy that forward adaptation exhibits lower LSD compared to self-adaptation at all bit rates.

Although objective evaluations provide valuable insights, they are not always well correlated with perception. Therefore, a subjective evaluation was conducted, consisting in a pairwise comparison between two 64 kb/s PCM codecs with unquantized forward- and self-adaptive pre- and de-emphasis filters. Both used $\gamma=0.7$. Ten speech files (five male and five female) were presented to eleven experienced subjects aged 30-55. The test did not include any training sessions. Each subject was given three options (A, B, or indifferent).
The audio setup consisted of a computer, a Focusrite Scarlett 2i2 audio interface, and Beyerdynamic DT 700 Pro headphones. A total of 110 votes were collected. Table \ref{tab:table1} presents the percentage of preferences. The ``Prob.'' line in the table indicates the probability that the observed preferences are solely due to random effects (sign test). These results reveal no statistically significant difference in perceived audio quality between forward-adaptive and self-adaptive pre- and de-emphasis.
 
\begin{table}[t!]
\centering
\caption{Subjective evaluation results}
\label{tab:table1}
\resizebox{0.6\columnwidth}{!}{%
\begin{tabular}{|c|c|}
\hline
\scriptsize Preference & \scriptsize Percentage of votes \\ \hline
\scriptsize SA         & \scriptsize 34.50              \\ \hline
\scriptsize FW         & \scriptsize 27.30              \\ \hline
\scriptsize SA = FW      & \scriptsize 38.20               \\ \hline
\scriptsize Prob.      & \scriptsize 0.19            \\ \hline
\end{tabular}%
}
\end{table}

\section{Conclusion}

This paper introduced a novel approach to adapt first-order pre- and de-emphasis filters. These filters are often applied to speech and audio signals before encoding and after decoding to simplify fixed-point implementation, facilitate LPC analysis, and improve subjective quality.

Evaluation results showed that the proposed self-adaptation approach brings similar subjective improvements to unquantized forward adaptation. However, the proposed approach does not consume any bit rate. Therefore, it does not require any bitstream modification and can be used in conjunction with any existing codec for a pre- and post-processing that is independent of the codec. It is interesting to note that forward adaptation normally requires quantizing the pre-emphasis coefficient, which in addition to costing bit rate can reduce subjective performance. Compared to backward adaptation, the proposed self-adaptation approach does not introduce any lag between filter and signal. It is also less complex because the decoded signal is not needed at the encoder.

The implementation that we tested introduces a small additional algorithmic delay because the analysis at the decoder requires some look-ahead (10 ms). But this can be optimized, for example by using asymmetric analysis windows \cite{chu2003window}. Also, self-adaptation does not guarantee that the pre- and de-emphasis coefficients will be identical, but subjective evaluations have shown that this is not an issue. Overall, the proposed zero-bit self-adaptive approach is a valuable alternative to the classical forward and backward adaptations.

\section{Future work}

While first-order self-adaptive pre- and de-emphasis filters proved effective, trying to extend this approach to higher orders resulted in a complex system of equations, making it challenging to find a general closed-form solution. Our objective now is to develop a comprehensive mathematical framework for higher-order filters for improved performance and broader applicability.

\vspace*{0.2cm}

{\footnotesize 
\bibliographystyle{unsrt}
\bibliography{Main}
}

\end{document}